\documentstyle[sprocl,epsfig]{article}

\def\be{\begin{equation}}
\def\ee{\end{equation}}
\def\bea{\begin{eqnarray}}
\def\eea{\end{eqnarray}}
\begin{document}

\title{ON THE NATURE OF HIDDEN BROAD LINE REGIONS IN SEYFERT 2 GALAXIES}

\author{A. MARTOCCHIA, G. MATT}

\address{Dipartimento di Fisica, Universit\`a degli Studi "Roma Tre"\\ 
Via della Vasca Navale, 84 -- I-00146 ROMA -- ITALY \\E-mail: 
martocchia@fis.uniroma3.it} 

%\author{A. N. OTHER}
%\address{Department of Physics, Theoretical Physics, 1 Keble Road,\\
%Oxford OX1 3NP, England\\E-mail: other@tp.ox.uk}

\maketitle\abstracts{We consider the sample of Tran (2001) 
to study the X-ray properties of Seyfert 2 galaxies 
with and without polarized broad lines. }

\section{Introduction }
%\subsection{Producing the Hard Copy}
\label{sec:intro}
Within the class of Seyfert 2 galaxies, two populations are 
observed:
in one, polarized broad emission lines are detected in the optical band, 
yielding evidence of {\it ``hidden broad-lines regions''} (HBLRs); 
in the spectra of the other class, such lines are not detected 
(non-HBLR Sy-2 galaxies). 
The question arises, whether the two classes are intrinsically the same
or not, that is, whether HBLRs are always there, even if sometimes - in
the case of non-HBLR Sy-2's - they are not seen, due e.g to obscuration.
Tran (2001), on the basis of the CfA and 
12 $\mu$m samples of Sy-2's, claimed that HBLR galaxies display
distinctly higher radio power relative to their infrared power, 
and hotter dust temperature. Since the average obscuration appears 
to be indistinguishable between the two classes, he proposed that
HBLRs galaxies harbor an energetic, hidden Sy-1 nucleus 
(where broad lines are produced), while non-HBLRs galaxies do not.
\cite{tr}

We went through Tran's sample, searching in the literature
for measurements of the intrinsic X-ray (2-10 keV) and OIII (5007 \AA)
luminosities ($L_X$, $L_{OIII}$), together with the column density 
($n_{\rm H}$), in order to test Tran's hypothesis of an intrinsic 
difference between the two classes. 

%\noindent
%{\sf http://www.wspc.com.sg/others/style\_files/proceedings/proceedings\_style\_files.html} }
%\noindent 
%{\em readme$\_$209.txt} --- the preliminary guide.
%\noindent 
%{\em sprocl.sty} --- the style file that provides the 
%higher level latex commands for the proceedings. Don't 
%change these parameters.
%\noindent 
%{\em sprocl.tex} --- the main text. 

\section{The sample }

We had to exclude many Sy-2 galaxies from the sample, for which: 
{\it either}
no X-ray observations have been performed with ``modern''
satellites (starting with {\it ASCA}), 
{\it or}
no estimates of the OIII flux could be found, 
{\it or}
any estimate of the intrinsic (nucleus) luminosity is
unreliable due to Compton-thickness ($n_{\rm H} > 10^{24}$ cm$^{-2}$).
We also considered additional galaxies, not belonging to
Tran's ensemble: based on the same criteria, we could finally 
add only the HBLR galaxy MKN 1210 in the sample. The final sample 
consists of 10 HBLR and 6 non-HBLR objects. 
The values of the luminosities are corrected for absorption.
All values listed in the tables have been taken from the best-available
observations. In some cases, the data have been re-analyzed.

\begin{table}[t]
%\vspace{0.2cm}
\begin{center}
\footnotesize

\begin{tabular}{|c|cccc|}\hline
Source name & 
$L_X$ & 
$L_{OIII}$ &
$n_{\rm H}$ ($\times 10^{22}$ cm$^{-2}$) & 
References \cr
\hline
\hline
& & & & \cr
IRAS 05189-2524 & 26.96 & 3.67 & 4.9   
& [2] \cr
IRAS 22017+0319 & 32.7 & 3.6$^{*}$  & 5.41
& [9],BOA$^{\#}$,[3] \cr
IC 5063 & 9.43 & 1.1  & 24    
& [2] \cr
MCG-3-34-64 & 4.3 & 2.68 & 79 
& [2] \cr
MKN 348 (NGC 262) &        12.19 & 0.97 & 10.6  
& [2] \cr
MKN 463E (UGC 8850) &      8.0   & 7.78 & 16    
& [2], [1] \cr
NGC 4388 &                7.44   & 0.65 & 42    
& [2] \cr
NGC 5506 &                10.07  & 0.56 & 3.4   
& [2], [8] \cr
NGC 6552 &                3.6    & 1.64 & 60    
& [2] \cr
UGC 4203 (MKN 1210) &    19     & 2.14 & 21.4  
& [2], [5] \cr
& & & & \cr
\hline
\end{tabular}

%\vspace{0.2in}
\caption{{\it Above:} Galaxies with HBLRs. The
luminosities are in units of $10^{42}$ erg s$^{-1}$.
A (*) near a $L_{OIII}$ value means that from the
literature it is not clear whether 
a correction for intrinsic absorption 
has been made or not. A (\#) marks values obtained through newly performed 
data analysis, using the standard XSPEC package. {\it BOA} stands for 
the online {\it BeppoSAX} data archive, accessible at 
{\sf http://www.asdc.asi.it/bepposax/archive\_browser.html},
while {\it AOA} stands for the {\it ASCA} online archive 
{\it Tartarus}, at {\sf http://tartarus.gsfc.nasa.gov/}.
{\it Below:} The non-HBLRs part of the sample. Units and conventions
are the same like above.  \label{table2}}
%\vspace{0.2in}

\begin{tabular}{|c|cccc|}
\hline
Source name & 
$L_X$ & 
$L_{OIII}$ &
$n_{\rm H}$ ($\times 10^{22}$ cm$^{-2}$) & 
References \cr
\hline
\hline
& & & & \cr
IRAS 00198-7926 & 2.04 & 1.55$^{*}$ & 0.26
& AOA$^{\#}$,[3] \cr
M51 (NGC 5194) & 0.008 & 0.013 & 75 
& [2] \cr
MKN 266 (NGC 5256) & 0.5 & 0.11$^{*}$ & 1.6
& BOA,[7],[3] \cr
NGC 3079 & 0.021 &   0.031 &   1.6 
& [2] \cr
NGC 4941 & 0.1   &   0.11  &   45  
& [2] \cr
%NGC 7172 & 3.94  &   0.007 &   9   & \cr
NGC 7582 (Grus Quartet) & 1.83 & 0.3 & 12.4 
& [2] \cr
& & & & \cr
\hline
\end{tabular}

\end{center}
\end{table}

\begin{figure}[h]
%\begin{tabular}{c}
\vskip -2 true cm
%\centerline{
\epsfig{figure=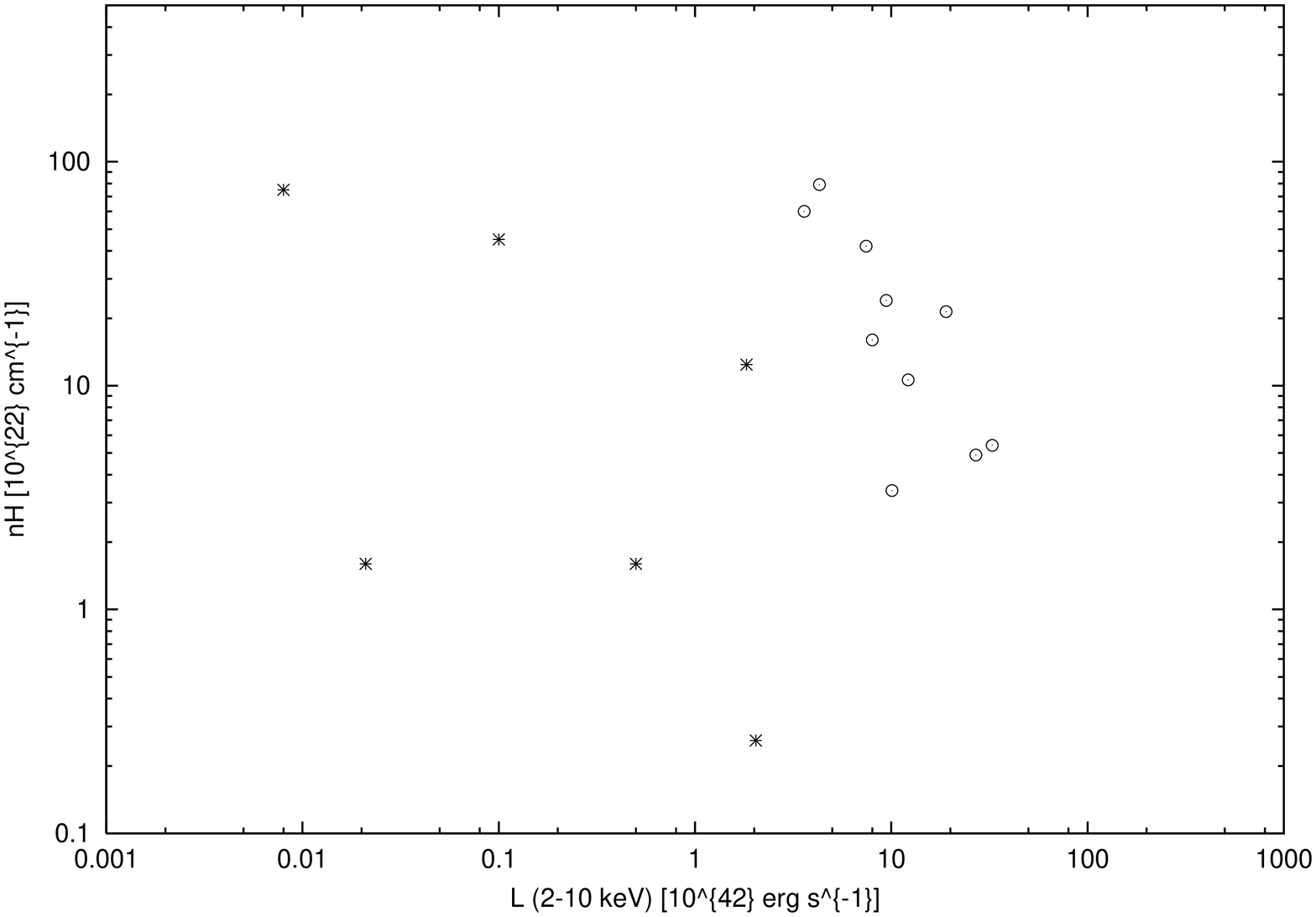, height=8cm, width=13cm,
angle=-90}     \\
%\vskip -0.3 true cm
%\centerline{
\epsfig{figure=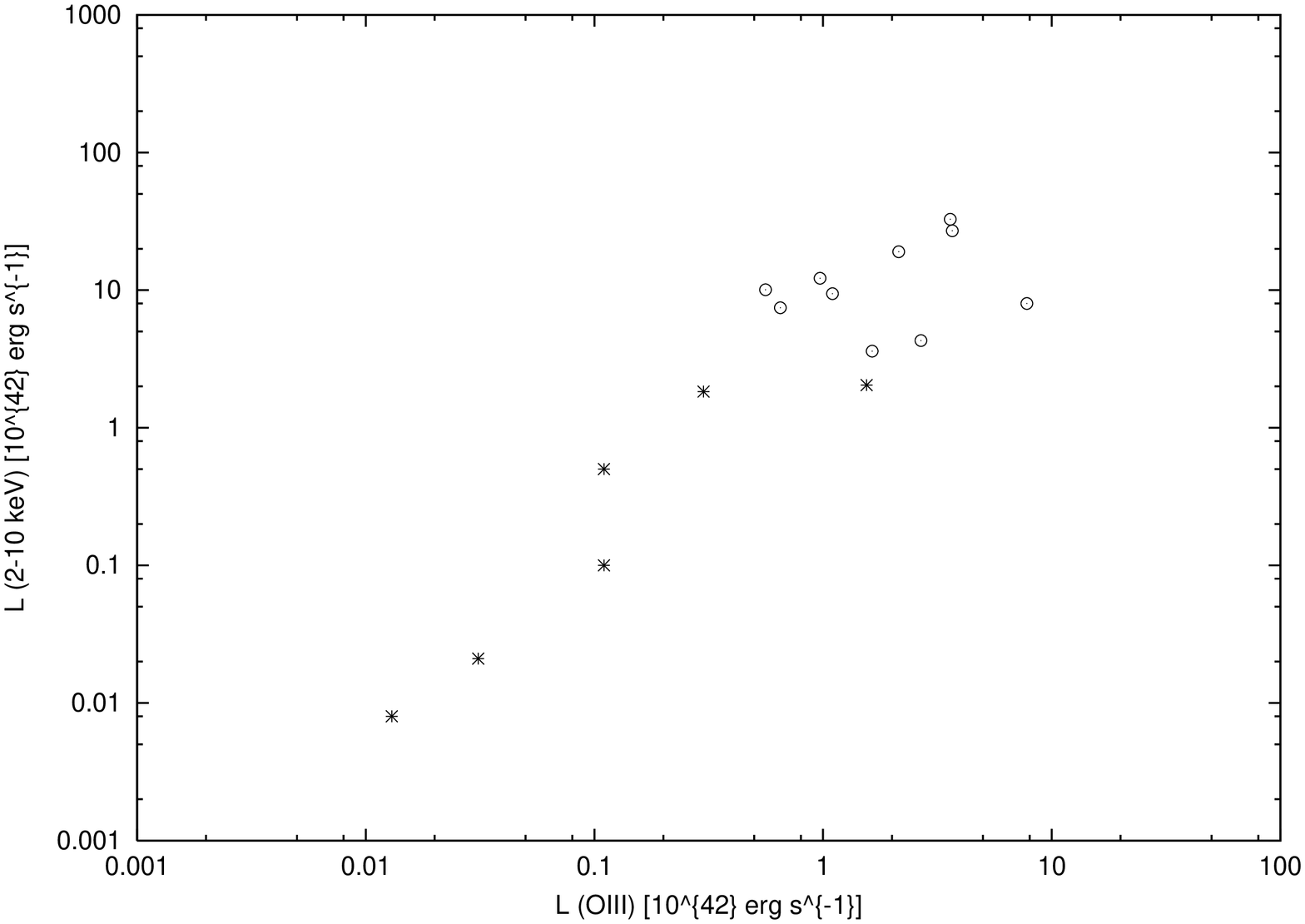, height=8cm, width=13cm,
angle=-90}
\caption{{\it Above:} Plot $n_{\rm H}$ versus $L_X$. 
{\it Below:} Plot $L_{OIII}$ versus $L_X$. In the plots, the simbol 
(*) is used for non-HBLR 
galaxies, ($\odot$) for objects which do show polarized broad lines. }
%\end{tabular}
\end{figure}

\section{Interpretation }
\label{sec:notes} 

%The sources varied in luminosity (e.g. M51) and/or 
%Compton thickness (e.g. MKN 1210, \cite{gu}) by 
%orders of magnitude from one observation to the other, 
%sometimes even ``disappeared'' (e.g. IRAS 00198-7926 is not 
%detected with {\it BeppoSAX}). Moreover,
%since many of Tran's sources have been observed with old-generation 
%satellites only (up to {\it Ginga}), 
While a systematic search with a large, 
complete and unbiased sample of Sy-2 galaxies is needed in order to
provide a more homogeneous and reliable set of data\footnote{After
this contribution was presented at the Conference, we became aware of
an indipendent study by Gu \& Huang (2002), in which a bigger compilation
of sources is given, and whose results are consistent with ours.\cite{gh} },
our preliminary statistical analysis seems to 
strengthen Tran's claim, since, as clearly shown in the plots,
\begin{itemize}
\item while the average extinction is indistinguishable between the two
classes (see also Tran\cite{tr}), the intrinsic 2-10 keV luminosity results 
to be sistematically smaller in non-HBLR Sy-2's, with a ``critical'' 
threshold around $\sim 5 \times 10^{42}$ erg s$^{-1}$;
\item galaxies without polarized broad lines appear to have fainter 
nuclei both in X and optical (see plot); in general, the X-ray vs. 
optical flux dependence is more or less a linear one.
\end{itemize}

How to explain the intrinsic difference? 
Tran (2001) claimed that HBLR galaxies are those which actually do
possess nuclei of Seyfert 1 type, while non-HBLR galaxies would not, 
the latter being ``powered'' mainly by starburst.\cite{tr} If this 
is the case,
broad lines regions simply {\it would not exist} in the cores of
those galaxies where polarized broad lines are not observed.
However, Seyfert 1 galaxies with $L_{2-10 {\rm keV}} < 10^{42}$ 
erg s$^{-1}$ {\it are} observed indeed, and there is no reason why
Sy-1 type nuclei (i.e. hidden broad line regions) with $L_{2-10 {\rm keV}}
< 10^{42}$ erg s$^{-1}$ should not be present in the cores of some 
Seyfert 2s, too. 

Even if, on the base of Tran's as well as our present study, 
a unifying model based {\it only} on inclination (such as the one by
Heisler et al.\cite{he}) 
seems to be ruled out, nevertheless one may suppose that 
{\it the intrinsic difference originates also in some property of the 
reflector}. One possible explanation is, 
that in the so-called non-HBLR Sy-2 galaxies ($L_{2-10 {\rm keV}} <
10^{42}$ erg s$^{-1}$) the irradiation from the nucleus is not strong 
enough to allow sublimation of dust grains and ionization of the more 
distant clouds (those located outside the torus). Thus, the broad lines 
cannot be reflected towards us, and are rather absorbed by the clouds 
themselves.
Conversely, in the so-called HBLR Sy-2s, the irradiation 
efficiently ionizes the distant clouds too, so that we can observe
polarized (reflected) broad lines.

\end{document}